\documentstyle[prl,aps,preprint]{revtex}
\newcommand{\wh}[1]{\widehat{#1}}
\newcommand{\dslash}[1]{#1 {\mkern-9mu} / }
\begin{document}
\setcounter{page}{1}
\def\footnoterule{\kern-3pt \hrule width\hsize \kern3pt}
\tighten
\title{Elimination of the vacuum instability for finite nuclei \\
in the relativistic $\sigma$-$\omega$ model}

\author{J. Caro${}^{1,2}$, E. Ruiz Arriola${}^1$ and L.L. Salcedo${}^{1,2}$}
\address{
{~} \\
${}^1$Departamento de F\'{\i}sica Moderna \\
Universidad de Granada \\
E-18071 Granada, Spain \\
{~} \\
${}^2$Center for Theoretical Physics \\
Laboratory for Nuclear Science \\
and Department of Physics \\
Massachusetts Institute of Technology \\
Cambridge, Massachusetts 02139, U.S.A. \\
{~}}


\date{MIT-CTP-2563,~  {~~~~~} September 1996}
\maketitle

\thispagestyle{empty}

\begin{abstract}

The $\sigma$-$\omega$ model of nuclei is studied at leading order in
the $1/N$ expansion thereby introducing the self consistent Hartree
approximation, the Dirac sea corrections and the one fermion loop
meson self energies in a unified way. For simplicity, the Dirac sea is
further treated within a semiclassical expansion to all orders. The
well-known Landau pole vacuum instability appearing in this kind of
theories is removed by means of a scheme recently proposed in this
context. The effect of such removal on the low momentum effective
parameters of the model, relevant to describe nuclear matter and
finite nuclei, is analyzed. The one fermion loop meson self energies
are found to have a sizeable contribution to these parameters.
However, such contribution turns out to come mostly from the Landau
poles and is thus spurious. We conclude that the fermionic loop can
only be introduced consistently in the $\sigma$-$\omega$ nuclear model
if the Landau pole problem is dealt with properly.

\end{abstract}



\eject

\section{Introduction}
\label{I}

The relativistic approach to nuclear physics has attracted much
attention. From a theoretical point of view, it allows to implement,
in principle, the important requirements of relativity, unitarity,
causality and renormalizability~\cite{Wa74}. From the phenomenological
side, it has also been successful in reproducing a large body of
experimental data~\cite{Wa74,Ho81,Se86,Re89,Se92}.  In the context of
finite nuclei a large amount of work has been done at the Hartree
level but considering only the positive energy single particle nucleon
states. The Dirac sea has also been studied since it is required to
preserve the unitarity of the theory. Actually, Dirac sea corrections
have been found to be non negligible using a semiclassical expansion
which, if computed to fourth order, seems to be quickly
convergent~\cite{Ca96a}. Therefore, it would appear that the overall
theoretical and phenomenological picture suggested by the relativistic
approach is rather reliable.

However, it has been known since ten years that such a description is
internally inconsistent. The vacuum of the theory is unstable due to
the existence of tachyonic poles in the meson propagators at high
Euclidean momenta~\cite{Pe87}.  Alternatively, a translationally
invariant mean field vacuum does not correspond to a minimum; the
Dirac sea vacuum energy can be lowered by allowing small size mean
field solutions~\cite{Co87}.  Being a short distance instability it
does not show up for finite nuclei at the one fermion loop level and
within a semiclassical expansion (which is an asymptotic large size
expansion).  For the same reason, it does not appear either in the
study of nuclear matter if translational invariance is imposed as a
constraint.  However, the instability sets in either in an exact mean
field valence plus sea (i.e., one fermion loop) calculation for finite
nuclei or in the determination of the correlation energy for nuclear
matter (i.e., one fermion loop plus a boson loop). Unlike quantum
electrodynamics, where the instability takes place far beyond its
domain of applicability, in quantum hadrodynamics it occurs at the
length scale of 0.2~fm that is comparable to the nucleon size and
mass.  Therefore, the existence of the instability contradicts the
original motivation that lead to the introduction of the field
theoretical model itself. In such a situation several possibilities
arise.  Firstly, one may argue that the model is defined only as an
effective theory, subjected to inherent limitations regarding the
Dirac sea.  Namely, the sea may at best be handled semiclassically,
hence reducing the scope of applicability of the model.  This
interpretation is intellectually unsatisfactory since the
semiclassical treatment would be an approximation to an inexistent
mean field description.  Alternatively, and taking into account the
phenomenological success of the model, one may take more seriously the
spirit of the original proposal~\cite{Wa74}, namely, to use specific
renormalizable Lagrangians where the basic degrees of freedom are
represented by nucleon and meson fields.  Such a path has been
explored in a series of papers~\cite{Ta90,Ta91,TB92} inspired by the
early work of Redmond and Bogolyubov on non asymptotically free
theories~\cite{Re58,Bo61}. The key feature of this kind of theories is
that they are only defined in a perturbative sense.  According to the
latter authors, it is possible to supplement the theory with a
prescription based on an exact fulfillment of the K\"all\'en-Lehmann
representation of the two point Green's functions.  The interesting
aspect of this proposal is that the Landau poles are removed in such a
way that the perturbative content of the theory remains unchanged. In
particular, this guarantees that the perturbative renormalizability is
preserved. It is, however, not clear whether this result can be
generalized to three and higher point Green's functions in order to
end up with a completely well-behaved field theory. Although the
prescription to eliminate the ghosts may seem to be ad hoc, it
certainly agrees more with the original proposal and provides a
workable calculational scheme.

The above mentioned prescription has already been used in the context
of nuclear physics. In ref.~\cite{Lo80}, it was applied to ghost
removal in the $\sigma$ exchange in the $NN$ potential. More recently, it
has been explored to study the correlation energy in nuclear matter in
the $\sigma$-$\omega$ model~\cite{TB92} and also in the evaluation of
response functions within a local density approximation~\cite{Ta90}.
Although this model is rather simple, it embodies the essential field
theoretical aspects of the problem while still providing a reasonable
phenomenological description. We will use the $\sigma$-$\omega$ model
in the present work, to estimate the binding energy of finite nuclei
within a self-consistent mean field description, including the effects
due to the Dirac sea, after explicit elimination of the ghosts. An
exact mean field calculation, both for the valence and sea, does make
sense in the absence of a vacuum instability but in practice it
becomes a technically cumbersome problem. This is due to the presence
of a considerable number of negative energy bound states in addition
to the continuum states\cite{Se92}. Therefore, it seems advisable to
use a simpler computational scheme to obtain a numerical estimate.
This will allow us to see whether or not the elimination of the ghosts
induces dramatic changes in the already satisfactory description of
nuclear properties. In this work we choose to keep the full Hartree
equations for the valence part but employ a semiclassical
approximation for the Dirac sea. This is in fact the standard
procedure~\cite{Se86,Re89,Se92}. As already mentioned, and discussed
in previous work~\cite{Ca96a}, this expansion converges rather quickly
and therefore might be reliably used to estimate the sea energy up to
possible corrections due to shell effects.

The paper is organized as follows. In section~\ref{II} we present the
$\sigma$-$\omega$ model of nuclei in the $1/N$ leading approximation,
the semiclassical treatment of the Dirac sea, the renormalization
prescriptions and the different parameter fixing schemes that we will
consider. In section~\ref{III} we discuss the vacuum instability
problem of the model and Redmond's proposal. We also study the
implications of the ghost subtraction on the low momentum effective
parameters. In section~\ref{IV} we present our numerical results for
the parameters as well as binding energies and mean quadratic charge
radii of some closed-shell nuclei. Our conclusions are presented in
section~\ref{V}. Explicit expressions for the zero momentum
renormalized meson self energies and related formulas are given in the
appendix.

\section{$\sigma$-$\omega$ model of nuclei}
\label{II}

In this section we revise the $\sigma$-$\omega$ model description of
finite nuclei disregarding throughout the instability problem; this
will be considered in the next section. The Dirac sea corrections are
included at the semiclassical level and renormalization issues as well
as the various ways of fixing the parameters of the model are also
discussed here.

\subsection{Field theoretical model}

Our starting point is the Lagrangian density of the $\sigma$-$\omega$
model~\cite{Wa74,Se86,Re89,Se92} given by
\begin{eqnarray}
{\cal L}(x) &=& \overline\Psi(x) \left[ \gamma_\mu ( i \partial^\mu - g_v
V^\mu(x)) - (M - g_s \phi(x)) \right] \Psi(x) 
+ {1\over 2}\, (\partial_\mu \phi(x)
\partial^\mu \phi(x) - m_s^2 \, \phi^2(x)) \nonumber \\
& & - {1\over 4} \, F_{\mu \nu}(x) F^{\mu \nu}(x)
+ {1\over 2} \, m_v^2 V_\mu(x) V^\mu(x) + \delta{\cal L}(x)\,.
\label{lagrangian}
\end{eqnarray}
$\Psi(x)$ is the isospinor nucleon field, $\phi(x)$ the scalar field,
$V_\mu(x)$ the $\omega$-meson field and $F_{\mu\nu} =\partial_\mu
V_\nu-\partial_\nu V_\mu$. In the former expression the necessary
counterterms required by renormalization are accounted for by the
extra Lagrangian term $\delta{\cal L}(x)$ (including meson
self-couplings).

Including Dirac sea corrections requires to take care of
renormalization issues. The best way of doing this in the present
context is to use an effective action formalism. Further we have to
specify the approximation scheme. The effective action will be
computed at lowest order in the $1/N$ expansion, $N$ being the number
of nucleon species (with $g_s$ and $g_v$ of order $1/\sqrt{N}$), that
is, up to one fermion loop and tree level for bosons~\cite{TB92}. This
corresponds to the Hartree approximation for fermions including the
Dirac sea~\cite{NO88}.

In principle, the full effective action would have to be computed by
introducing bosonic and fermionic sources. However,
since we will consider only stationary situations, we do not
need to introduce fermionic sources. Instead, we will
proceed as usual by integrating out exactly the fermionic degrees of
freedom. This gives directly the bosonic effective action at leading
order in the $1/N$ expansion:
\begin{equation}
\Gamma[\phi,V]= \Gamma_B[\phi,V]+\Gamma_F[\phi,V] \,,
\end{equation}
where
\begin{equation}
\Gamma_B[\phi,V]=\int\left({1\over 2}\, (\partial_\mu \phi
\partial^\mu \phi - m_s^2 \, \phi^2) - {1\over 4} \, F_{\mu \nu} F^{\mu \nu}
+ {1\over 2} \, m_v^2 V_\mu V^\mu \right) d^4x \,,
\label{GammaB}
\end{equation}
and
\begin{equation}
\Gamma_F[\phi,V]= -i\log {\rm Det}\left[ \gamma_\mu ( i \partial^\mu - g_v
V^\mu) - (M - g_s \phi) \right] +\int\delta{\cal L}(x)d^4x \,.
\end{equation}
The fermionic determinant can be computed perturbatively, by adding up
the one-fermion loop amputated graphs with any number of bosonic legs,
using a gradient expansion or by any other technique. The ultraviolet
divergences are to be canceled with the counterterms by using any
renormalization scheme; all of them give the same result after fitting
to physical observables.

The effective action so obtained is uniquely defined and completely
finite. However, there still remains the freedom to choose different
variables to express it. We will work with fields renormalized at zero
momentum. That is, the bosonic fields $\phi(x)$ and $V_\mu(x)$ are
normalized so that their kinetic energy term is the canonical
one. This is the choice shown above in $\Gamma_B[\phi,V]$. Other usual
choice is the on-shell one, namely, to rescale the fields so that the
residue of the propagator at the meson pole is unity. Note that the
Lagrangian mass parameters $m_s$ and $m_v$ do not correspond to the
physical masses (which will be denoted $m_\sigma$ and $m_\omega$ in
what follows) since the latter are defined as the position of the
poles in the corresponding propagators. The difference comes from the
fermion loop self energy in $\Gamma_F[\phi,V]$ that contains terms
quadratic in the boson fields with higher order gradients.

Let us turn now to the fermionic contribution, $\Gamma_F[\phi,V]$. We
will consider nuclear ground states of spherical nuclei, therefore the
space-like components of the $\omega$-meson field vanish \cite{Se92}
and the remaining fields, $\phi(x)$ and $V_0(x)$ are stationary.  As
it is well-known, for stationary fields the fermionic energy, i.e.,
minus the action $\Gamma_F[\phi,V]$ per unit time, can be formally written
as the sum of single particle energies of the fermion moving in the
bosonic background~\cite{NO88},
\begin{equation}
E_F[\phi,V_0] = \sum_n E_n \,,
\end{equation}
and
\begin{equation}
\left[ -i {\bf \alpha} \cdot \nabla + g_vV_0(x) + \beta (
M-g_s\phi(x)) \right] \psi_n(x) = E_n \,\psi_n(x)\,.
\label{Dirac-eq}
\end{equation}
Note that what we have called the fermionic energy contains not only
the fermionic kinetic energy, but also the potential energy coming
from the interaction with the bosons.

The orbitals, and thus the fermionic energy, can be divided into
valence and sea, i.e., positive and negative energy orbitals. In
realistic cases there is a gap in the spectrum which makes such a
separation a natural one. The valence energy is therefore given by
\begin{equation}
E_F^{\rm val}[\phi,V] = \sum_n E_n^{\rm val}\,.
\end{equation}
On the other hand, the sea energy is ultraviolet divergent and
requires the renormalization mentioned above~\cite{Se86}. The (at zero
momentum) renormalized sea energy is known in a gradient or
semiclassical expansion up to fourth order and is given by~\cite{Ca96a}
\begin{eqnarray}
E^{\rm sea}_0 & = & -{\gamma\over 16\pi^2} M^4 \int d^3x \,
\Biggl\{ \Biggr.
\left({\Phi\over M}\right)^4 \log {\Phi\over M}
+ {g_s\phi\over M} - {7\over 2} \left({g_s \phi\over M}\right)^2
+ {13\over 3} \left({g_s \phi\over M}\right)^3 - {25\over 12}
\left({g_s \phi\over M}\right)^4 \Biggl. \Biggr\} \nonumber\\ 
E^{\rm sea}_2 & = & {\gamma \over 16 \pi^2} \int d^3 x \,
\Biggl\{ {2 \over 3} \log{\Phi\over M} (\nabla V)^2
- \log{\Phi\over M} (\nabla \Phi)^2 \Biggr\} \nonumber\\
E^{\rm sea}_4 & = & {\gamma\over 5760 \pi^2} \int d^3 x \,
\Biggl\{ \Biggr. -11\,\Phi^{-4} (\nabla \Phi)^4
- 22\,\Phi^{-4} (\nabla V)^2(\nabla \Phi)^2
+ 44 \, \Phi^{-4} \bigl( (\nabla_i \Phi) (\nabla_i V) \bigr)^2
\nonumber\\
 & & \quad - 44 \, \Phi^{-3} \bigl( (\nabla_i \Phi) (\nabla_i V)
\bigr) (\nabla^2 V)
- 8 \, \Phi^{-4} (\nabla V)^4 + 22 \, \Phi^{-3} (\nabla^2 \Phi)
(\nabla \Phi)^2
\nonumber\\
 & & \quad
+ 14 \, \Phi^{-3} (\nabla V)^2 (\nabla^2 \Phi)
- 18 \, \Phi^{-2} (\nabla^2 \Phi)^2 + 24 \, \Phi^{-2} (\nabla^2 V)^2
\Biggl. \Biggr\}\,.
\label{Esea}
\end{eqnarray}
Here, $V=g_vV_0$, $\Phi=M-g_s\phi$ and $\gamma$ is the spin and
isospin degeneracy of the nucleon, i.e., $2N$ if there are $N$ nucleon
species (in the real world $N=2$). The sea energy is obtained by
adding up the terms above. The fourth and higher order terms are
ultraviolet finite as follows from dimensional counting. The first two
terms, being renormalized at zero momentum, do not contain operators
with dimension four or less, such as $\phi^2$, $\phi^4$, or $(\nabla
V)^2$, since they are already accounted for in the bosonic term
$\Gamma_B[\phi,V]$. Note that the theory has been renormalized so that
there are no three- or four-point bosonic interactions in the
effective action at zero momentum~\cite{Se86}.

By definition, the true value of the classical fields 
(i.e., the value in the absence of external
sources) is to be found by minimization of the
effective action or, in the stationary case, of the energy
\begin{equation}
E[\phi,V] = E_B[\phi,V]+E_F^{\rm val}[\phi,V] + E_F^{\rm sea}[\phi,V] \,.
\end{equation}
Such minimization yields the equations of motion for the bosonic
fields:
\begin{eqnarray}
(\nabla^2-m_s^2)\phi(x) &=& -g_s\left[\rho_s^{\rm val}(x)+
\rho_s^{\rm sea}(x)\right] \,, \nonumber \\
(\nabla^2-m_v^2)V_0(x) &=& -g_v\left[\rho^{\rm val}(x)+
\rho^{\rm sea}(x) \right] \,.
\label{Poisson-eq}
\end{eqnarray}
Here, $\rho_s(x)=\langle\overline\Psi(x)\Psi(x)\rangle$
is the scalar density and $\rho(x)=\langle\Psi^\dagger(x)
\Psi(x)\rangle$ the baryonic one:
\begin{eqnarray}
\rho_s^{\rm val~(sea)}(x) &=& -\frac{1}{g_s}\frac{\delta E_F^{\rm
val~(sea)}}{\delta \phi(x)}\,, \nonumber \\
\rho^{\rm val~(sea)}(x) &=& +\frac{1}{g_v}\frac{\delta E_F^{\rm
val~(sea)}}{\delta V_0(x)} \,.
\label{densities}
\end{eqnarray}
The set of bosonic and fermionic equations, eqs.~(\ref{Poisson-eq})
and (\ref{Dirac-eq}) respectively, are to be solved self-consistently.
Let us remark that treating the fermionic sea energy using a gradient
or semiclassical expansion is a further approximation on top of the
mean field approximation since it neglects possible shell effects in
the Dirac sea. However, a direct solution of the mean field equations
including renormalization of the sum of single-particle energies would
not give a physically acceptable solution due to the presence of
Landau ghosts. They will be considered in the next section.

At this point it is appropriate to make some comments on
renormalization. As we have said, one can choose different
normalizations for the mesonic fields and there are also several sets
of mesonic masses, namely, on-shell and at zero momentum. If one were
to write the mesonic equations of motion directly, by similarity with
a classical treatment, there would be an ambiguity as to which set
should be used. The effective action treatment makes it clear that the
mesonic field and masses are those at zero momentum. On the other
hand, since we have not included bosonic loops, the fermionic
operators in the Lagrangian are not renormalized and there are no
proper vertex corrections. Thus the nucleon mass $M$, the nuclear
densities $\langle\Psi\overline\Psi\rangle$ and the combinations
$g_s\phi(x)$ and $g_v V_\mu(x)$ are fixed unambiguously in the
renormalized theory. The fermionic energy $E_F[\phi,V]$, the
potentials $\Phi(x)$ and $V(x)$ and the nucleon single particle
orbitals are all free from renormalization ambiguities at leading
order in $1/N$.

\subsection{Fixing of the parameters}

The $\sigma$-$\omega$ and related theories are effective models of
nuclear interaction, and hence their parameters are to be fixed to
experimental observables within the considered approximation. Several
procedures to perform the fixing can be found in the literature
\cite{Ho81,Re89,Se92}; the more sophisticated versions try to adjust,
by minimizing the appropriate $\chi^2$ function, as many experimental
values as possible through the whole nuclear table \cite{Re89}. These
methods are useful when the theory implements enough physical elements
to provide a good description of atomic nuclei. The particular model
we are dealing with can reproduce the main features of nuclear force,
such as saturation and the correct magic numbers; however it lacks
many of the important ingredients of nuclear interaction, namely
Coulomb interaction and $\rho$ and $\pi$ mesons. Therefore, we will
use the simple fixing scheme proposed in ref.~\cite{Ho81} for
this model.

Initially there are five free parameters: the nucleon mass ($M$), two
boson Lagrangian masses ($m_s$ and $m_v$) and the corresponding
coupling constants ($g_s$ and $g_v$). The five physical observables to
be reproduced are taken to be the physical nucleon mass, the physical
$\omega$-meson mass $m_\omega$, the saturation properties of nuclear
matter (binding energy per nucleon $B/A$ and Fermi momentum $k_F$) and
the mean quadratic charge radius of $^{40}$Ca. In our approximation,
the equation of state of nuclear matter at zero temperature, and hence
its saturation properties, depends only on the nucleon mass and on
$m_{s,v}$ and $g_{s,v}$ through the combinations~\cite{Se86}
\begin{equation}
C_s^2 = g_s^2 \frac{M^2}{m_s^2}\,, \qquad
C_v^2 = g_v^2 \frac{M^2}{m_v^2}\,.
\end{equation}

At this point, there still remain two parameters to be fixed, e.g.,
$m_v$ and $g_s$.  Now we implement the physical $\omega$-meson mass
constraint. From the expression of the $\omega$ propagator at the
leading $1/N$ approximation, we can obtain the value of the physical
$\omega$ pole as a function of the Lagrangian parameters $M$, $g_v$
and $m_v$ or more conveniently as a function of $M$, $C_v$ and $m_v$
(see appendix). Identifying the $\omega$ pole and the physical
$\omega$ mass, and given that $M$ and $C_v$ have already been fixed,
we obtain the value of $m_v$. Finally, the value of $g_s$ is adjusted
to fit the mean quadratic charge radius of $^{40}$Ca. We will refer to
this fixing procedure as the {\em $\omega$-shell scheme}: the name
stresses the correct association between the pole of the
$\omega$-meson propagator and the physical $\omega$ mass. The above
fixing procedure gives different values of $m_s$ and $g_s$ depending
on the order at which the Dirac sea energy is included in the
semiclassical expansion (see section~\ref{IV}).

Throughout the literature the standard fixing procedure when the Dirac
sea is included has been to give to the Lagrangian mass $m_v$ the
value of the physical $\omega$ mass \cite{Re89,Se92} (see, however,
refs.~\cite{Ta91,Ca96a}). Of course, this yields a wrong value for the
position of the $\omega$-meson propagator pole, which is
underestimated. We will refer to this procedure as the {\em naive
scheme}. Note that when the Dirac sea is not included at all, the
right viewpoint is to consider the theory at tree level, and the
$\omega$-shell and the naive schemes coincide.

\section{Landau instability subtraction}
\label{III}

As already mentioned, the $\sigma$-$\omega$ model, and more generally
any Lagrangian which couples bosons with fermions by means of a
Yukawa-like coupling, exhibits a vacuum instability~\cite{Pe87,Co87}.
This instability prevents the actual calculation of physical
quantities beyond the mean field valence approximation in a systematic
way. Recently, however, a proposal by Redmond~\cite{Re58} that
explicitly eliminates the Landau ghost has been implemented to
describe relativistic nuclear matter in a series of
papers~\cite{Ta90,Ta91,TB92}. The main features of such method are
contained already in the original papers and many details have also
been discussed. For the sake of clarity, we outline here the method as
applies to the calculation of Dirac sea effects for closed-shell
finite nuclei.

\subsection{Landau instability}

Since the Landau instability shows up already at zero nuclear density,
we will begin by considering the vacuum of the $\sigma$-$\omega$
theory. On a very general basis, namely, Poincar\'e invariance,
unitarity, causality and uniqueness of the vacuum state, one can show
that the two point Green's function (time ordered product) for a
scalar field admits the K\"all\'en-Lehmann representation~\cite{BD65}
\begin{eqnarray}
D(x'-x) = \int\,d\mu^2\rho(\mu^2)\,D_0(x'-x\, ;\mu^2)\,,
\label{KL}
\end{eqnarray}
where the full propagator in the vacuum is 
\begin{eqnarray}
D(x'-x) = -i \langle
0|T\phi(x')\phi(x)|0\rangle\,,
\end{eqnarray}
and the free propagator reads
\begin{eqnarray}
D_0(x'-x\,;\mu^2) = \int\,{d^4p\over (2\pi)^4}{ e^{-ip(x'-x)}\over
p^2-\mu^2+i\epsilon }\,.
\end{eqnarray}
The spectral density $\rho(\mu^2)$ is
defined as 
\begin{eqnarray}
\rho(q^2) = (2\pi)^3\sum_n\delta^4(p_n-q)|\langle
0|\phi(0)|n\rangle|^2\,.
\end{eqnarray}
It is non negative, Lorentz invariant and vanishes for space-like four
momentum $q$.

The K\"all\'en-Lehmann representability is a necessary condition for
any acceptable theory, yet it is violated by the $\sigma$-$\omega$
model when the meson propagators are approximated by their leading
$1/N$ term. It is not clear whether this failure is tied to the theory
itself or it is an artifact of the approximation ---it is well-known
that approximations to the full propagator do not necessarily preserve
the K\"all\'en-Lehmann representability---. The former possibility
would suppose a serious obstacle for the theory to be a reliable one.

In the above mentioned approximation, eq.~(\ref{KL}) still holds both for the
$\sigma$ and the $\omega$ cases (in the latter case with obvious
modification to account for the Lorentz structure)
but the spectral density gets modified to be
\begin{eqnarray}
\rho(\mu^2) = \rho^{\rm KL}(\mu^2) - R_G\delta(\mu^2+M_G^2)
\end{eqnarray}
where $\rho^{\rm KL}(\mu^2)$ is a physically acceptable spectral
density, satisfying the general requirements of a quantum field
theory. On the other hand, however, the extra term spoils these
general principles.  The residue $-R_G$ is negative, thus indicating
the appearance of a Landau ghost state which contradicts the usual
quantum mechanical probabilistic interpretation. Moreover, the delta
function is located at the space-like squared four momentum $-M_G^2$
indicating the occurrence of a tachyonic instability. As a
perturbative analysis shows, the dependence of $R_G$ and $M_G$ with
the fermion-meson coupling constant $g$ in the weak coupling regime is
$R_G\sim g^{-2}$ and $M_G^2 \sim 4M^2\exp(4\pi^2/g^2)$, with $M$ the
nucleon mass.  Therefore the perturbative content of $\rho(\mu^2)$ and
$\rho^{\rm KL}(\mu^2)$ is the same, i.e., both quantities coincide
order by order in a power series expansion of $g$ keeping $\mu^2$
fixed. This can also be seen in the propagator form of the previous
equation
\begin{eqnarray}
D(p) = D^{\rm KL}(p) - {R_G\over p^2+M_G^2}\,.
\label{Delta}
\end{eqnarray}
For fixed four momentum, the ghost term vanishes as
$\exp(-4\pi^2/g^2)$ when the coupling constant goes to zero. As noted
by Redmond~\cite{Re58}, it is therefore possible to modify the theory
by adding a suitable counterterm to the action that exactly cancels
the ghost term in the meson propagator without changing the
perturbative content of the theory. In this way the full meson
propagator becomes $D^{\rm KL}(p)$ which is physically acceptable and
free from vacuum instability at leading order in the $1/N$ expansion.

It is not presently known whether the stability problems of the
original $\sigma$-$\omega$ theory are intrinsic or due to the
approximation used, thus Redmond's procedure can be interpreted either
as a fundamental change of the theory or as a modification of the
approximation scheme. Although both interpretations use the
perturbative expansion as a constraint, it is not possible, at the
present stage, to decide between them.  It should be made quite clear
that in spite of the seemingly arbitrariness of the no-ghost
prescription, the original theory itself was ambiguous regarding its
non perturbative regime. In fact, being a non asymptotically free
theory, it is not obvious how to define it beyond finite order
perturbation theory. For the same reason, it is not Borel summable and
hence additional prescriptions are required to reconstruct the Green's
functions from perturbation theory to all orders. As an example, if
the nucleon self energy is computed at leading order in a $1/N$
expansion, the existence of the Landau ghost in the meson propagator
gives rise to a pole ambiguity. This is unlike physical time-like
poles, which can be properly handled by the customary $+i\epsilon$
rule, and thus an additional ad hoc prescription is needed. This
ambiguity reflects in turn in the Borel transform of the perturbative
series; the Borel transform presents a pole, known as renormalon in
the literature~\cite{Zi79}. In recovering the sum of the perturbative
series through inverse Borel transformation a prescription is then
needed, and Redmond's proposal provides a particular suitable way of
fixing such ambiguity. Nevertheless, it should be noted that even if
Redmond's prescription turns out to be justified, there still remains
the problem of how to extend it to the case of three- and more point
Green's functions, since the corresponding K\"all\'en-Lehmann
representations has been less studied.

\subsection{Instability subtraction}

To implement Redmond's prescription in detail we start with the
zero-momentum renormalized propagator in terms of the proper
self-energy for the scalar field (a similar construction can be
carried out for the vector field as well),
\begin{eqnarray}
D_s(p^2) = (p^2-m_s^2 - \Pi_s(p^2))^{-1}\,,
\end{eqnarray}
where the $m_s$ is the zero-momentum meson mass and the corresponding
renormalization conditions are $\Pi_s(0)= \Pi_s^\prime(0)=0$. The
explicit formulas for the scalar and vector meson self energies are
given in the appendix. Of course, $D_s(p^2)$ is just the inverse of
the quadratic part of the effective action $K_s(p^2)$. According to
the previous section, the propagator presents a tachyonic pole.  Since
the ghost subtraction is performed at the level of the two-point
Green's function, it is clear that the corresponding Lagrangian
counterterm must involve a quadratic operator in the mesonic fields.
The counterterm kernel $\Delta K_s(p^2)$ must be such that cancels the
ghost term in the propagator $D_s(p^2)$ in eq.~(\ref{Delta}). The
subtraction does not modify the position of the physical meson pole
nor its residue, but it will change the zero-momentum parameters and
also the off-shell behavior. Both features are relevant to nuclear
properties. This will be discussed further in the next section.

Straightforward calculation yields
\begin{eqnarray}
\Delta K_s(p^2) = 
-{1\over D_s(p^2)}{R_G^s\over R_G^s+(p^2+{M^s_G}^2)D_s(p^2)} \,.
\label{straightforward}
\end{eqnarray}
As stated, this expression vanishes as $\exp(-4\pi^2/g_s^2)$ for small
$g_s$ at fixed momentum. Therefore it is a genuine non perturbative
counterterm. It is also non local as it depends in a non polynomial
way on the momentum. In any case, it does not introduce new
ultraviolet divergences at the one fermion loop level. However, it is
not known whether the presence of this term spoils any general
principle of quantum field theory.

Proceeding in a similar way with the $\omega$-field $V_\mu(x)$, the
following change in the total original action is induced
\begin{eqnarray}
\Delta S = {1\over 2}\int{d^4p\over (2\pi)^4}\phi(-p)
\Delta K_s(p^2)\phi(p) - {1\over 2}\int{d^4p\over
(2\pi)^4}V_\mu(-p)\Delta K^{\mu\nu}_v(p^2)V_\nu(p) \,,
\label{Delta S}
\end{eqnarray}
where $\phi(p)$ and $V_\mu(p)$ are the Fourier transform of the scalar
and vector fields in coordinate space, $\phi(x)$ and $V_\mu(x)$
respectively. Note that at tree-level for bosons, as we are
considering throughout, this modification of the action is to be added
directly to the effective action ---in fact, this is the simplest way
to derive eq.~(\ref{straightforward})---.
Therefore, in the case of static fields, the total mean field energy
after ghost elimination reads
\begin{equation}
E= E_F^{\rm val} + E_F^{\rm sea} + E_B + \Delta E \,,
\end{equation}
where $E_F^{\rm val}$, $E_F^{\rm sea}$ and $E_B$ were given in
section~\ref{II} and
\begin{equation}
\Delta E[\phi,V] = {1\over 2}\int\,d^3x\phi(x)
\Delta K_s(\nabla^2)\phi(x) - {1\over 2}\int\,d^3x 
V_0(x)\Delta K^{00}_v(\nabla^2)V_0(x) \,.
\end{equation}
One can proceed by minimizing the mean field total energy as a
functional of the bosonic and fermionic fields. This yields the usual
set of Dirac equations for the fermions, eqs.~(\ref{Dirac-eq}) and
modifies the left-hand side of the bosonic eqs.~(\ref{Poisson-eq}) by
adding a linear non-local term. This will be our starting point to
study the effect of eliminating the ghosts in the description of
finite nuclei. We note that the instability is removed at the
Lagrangian level, i.e., the non-local counterterms are taken to be new
terms of the starting Lagrangian which is then used to describe the
vacuum, nuclear matter and finite nuclei. Therefore no new
prescriptions are needed in addition to Redmond's to specify how the
vacuum and the medium parts of the effective action are modified by
the removal of the ghosts.

So far, the new counterterms, although induced through the Yukawa
coupling with fermions, have been treated as purely bosonic terms.
Therefore, they do not contribute directly to bilinear fermionic
operators such as baryonic and scalar densities. An alternative
viewpoint would be to take them rather as fermionic terms, i.e., as a
(non-local and non-perturbative) redefinition of the fermionic
determinant.  The energy functional, and thus the mean field equations
and their solutions, coincide in the bosonic and fermionic
interpretations of the new term, but the baryonic densities and
related observables would differ, since they pick up a new
contribution given the corresponding formulas similar to
eqs.~(\ref{densities}). Ambiguities and redefinitions are ubiquitous
in quantum field theories, due to the well-known ultraviolet
divergences. However, in well-behaved theories the only freedom
allowed in the definition of the fermionic determinant comes from
adding counterterms which are local and polynomial in the
fields. Since the new counterterms induced by Redmond's method are not
of this form, we will not pursue such alternative point of view in
what follows. Nevertheless, a more compelling argument would be needed
to make a reliable choice between the two possibilities.

\subsection{Application to finite nuclei}

In this section we will take advantage of the smooth behavior of the
mesonic mean fields in coordinate space which allows us to apply a
derivative or low momentum expansion. The quality of the gradient
expansion can be tested a posteriori by a direct computation. The
practical implementation of this idea consists of treating the term
$\Delta S$ by expanding each of the kernels $\Delta K(p^2)$ in a power
series of the momentum squared around zero
\begin{equation}
\Delta K(p^2) = \sum_{n\ge 0}\Delta K_{2n}\, p^{2n}\,.
\end{equation}
The first two terms are given explicitly by
\begin{eqnarray}
\Delta K_0 & = & -\frac{m^4R_G}{M_G^2-m^2R_G} ,\nonumber\\
\Delta K_2 & = & \frac{m^2 R_G(m^2-m^2R_G+2M_G^2)}{(M_G^2-m^2R_G)^2}.
\end{eqnarray}
The explicit expressions of the tachyonic pole
parameters $M_G$ and $R_G$ for each meson can be found below.

Numerically, we have found that the fourth and higher orders in this
gradient expansion are negligible as compared to zeroth- and
second orders. In fact, in ref.~\cite{Ca96a} the same behavior was
found for the correction to the Dirac sea contribution to the binding
energy of a nucleus. As a result, even for light nuclei, $E_4^{\rm
sea}$ in eq.~(\ref{Esea}) can be safely neglected. Furthermore, it has
been shown~\cite{Ca96} that the fourth order term in the gradient
expansion of the valence energy, if treated semiclassically, is less
important than shell effects.  So, it seems to be a general rule that,
for the purpose of describing static nuclear properties, only the two
lowest order terms of a gradient expansion need to be considered. We
warn, however, that the convergence of the gradient or semiclassical
expansion is not the same as converging to the exact mean field
result, since there could be shell effects not accounted for by this
expansion at any finite order. Such effects, certainly exist in the
valence part~\cite{Ca96}. Even in a seemingly safe case as infinite
nuclear matter, where only the zeroth order has a non vanishing
contribution, something is left out by the gradient expansion since
the exact mean field solution does not exist due to the Landau ghost
instability (of course, the situation may change if the Landau pole is
removed). In other words, although a gradient expansion might appear
to be exact in the nuclear matter case, it hides the very existence of
the vacuum instability.

From the previous discussion it follows that the whole effect of the
ghost subtraction is represented by adding a term $\Delta S$ to the
effective action with same form as the bosonic part of the original
theory, $\Gamma_B[\phi,V]$ in eq.~(\ref{GammaB}). This amounts to a
modification of the zero-momentum parameters of the effective
action. The new zero-momentum scalar field (i.e., with canonical
kinetic energy), mass and coupling constant in terms of those of the
original theory are given by
\begin{eqnarray}
{\wh\phi}(x) &=& (1+\Delta K^s_2)^{1/2}\phi(x)\,, \nonumber \\
\wh{m}_s &=& \left(\frac{m_s^2-\Delta K^s_0}{1+\Delta
K^s_2}\right)^{1/2} \,, \\
\wh{g}_s &=& (1+\Delta K^s_2)^{-1/2}g_s \,. \nonumber
\end{eqnarray}
The new coupling constant is obtained recalling that $g_s\phi(x)$
should be invariant. Similar formulas hold for the vector meson.  With
these definitions (and keeping only $\Delta K_{s,v}(p^2)$ till second
order in $p^2$) one finds\footnote{Note that $E_{B,F}[~]$ refer to the
functionals (the same at both sides of the equations) and not to their
value as is also usual in physics literature.}
\begin{eqnarray}
E_B[\wh{\phi},\wh{V};\wh{m}_s,\wh{m}_v] &=&
E_B[\phi,V; m_s,m_v] + \Delta E[\phi,V; m_s,m_v] \,, \\
E_F[\wh{\phi},\wh{V};\wh{g}_s,\wh{g}_v] &=& 
E_F[\phi,V;g_s,g_v] \,. \nonumber
\end{eqnarray}
The bosonic equations for the new meson fields after ghost removal are
hence identical to those of the original theory using
\begin{eqnarray}
\wh{m}^2  &=&  m^2 \, M_G^2 \, \frac{M_G^2 - m^2\, R_G}{M_G^4 + m^4 \, R_G }
\,,\nonumber \\
\wh{g}^2  &=&  g^2 \, \frac{(M_G^2 - m^2\, R_G)^2}{M_G^4 + m^4 \, R_G } \,,
\label{mg}
\end{eqnarray}
as zero-momentum masses and coupling constants respectively. In the
limit of large ghost masses or vanishing ghost residues, the
reparameterization becomes trivial, as it should be.  Let us note that
although the zero-momentum parameters of the effective action
$\wh{m}_{s,v}$ and $\wh{g}_{s,v}$ are the relevant ones for nuclear
structure properties, the parameters $m_{s,v}$ and $g_{s,v}$ are the
(zero-momentum renormalized) Lagrangian parameters and they are also
needed, since they are those appearing in the Feynman rules in a
perturbative treatment of the model. Of course, both sets of parameters
coincide when the ghosts are not removed or if there were no ghosts in
the theory.

\begin{figure}
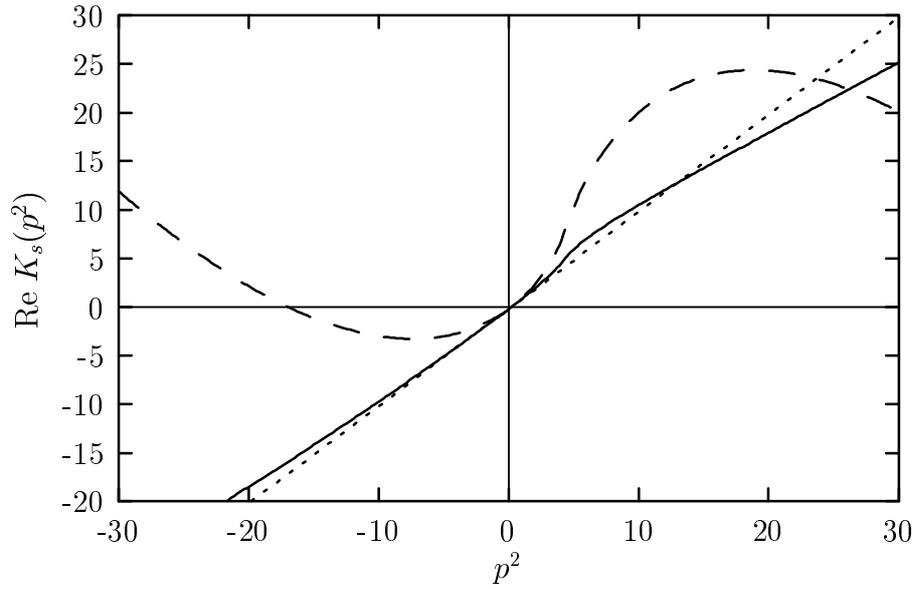

\caption{ Real part of the inverse scalar meson propagator $K_s(p^2)$
as a function of the squared four momentum using the no-ghost (sea 2nd)
set of parameters of table I. The dashed line represents the one-loop
result without ghost subtraction. The solid line is the result after
ghost elimination. The dotted line shows the free inverse
propagator. In all cases the slope at zero momentum is unity. Units
are in nucleon mass.}
\label{f-real}
\end{figure}

To finish this section we give explicitly the fourth order
coefficient in the gradient expansion of $\Delta E$, taking into account
the rescaling of the mesonic fields, namely,
\begin{equation}
\frac{\Delta K_4}{1+\Delta K_2}=
 -\frac{R_G (M_G^2 + m^2)^2}{
(M_G^4 +  m^4 \, R_G)\, (M_G^2 - m^2 \, R_G)}
- \frac{\gamma g^2}{\alpha \, \pi^2 \, M^2} \,
\frac{m^4 \, R_G^2 - 2 \, m^2 \, M_G^2 \, R_G }
{M_G^4 + m^4 \, R_G } \,,
\end{equation}
where $\alpha$ is $160$ for the scalar meson and $120$ for the vector
meson. As already stated, for typical mesonic profiles the
contribution of these fourth order terms are found to be numerically
negligible. Simple order of magnitude estimates show that squared
gradients are suppressed by a factor $(RM_G)^{-2}$, $R$ being the
nuclear radius, and therefore higher orders can also be
neglected. That the low momentum region is the one relevant to nuclear
physics can also be seen from the kernel $K_s(p^2)$, shown in
fig.~\ref{f-real}. From eq.~(\ref{Delta S}), this kernel is to be
compared with the function $\phi(p)$ that has a width of the order of
$R^{-1}$. It is clear from the figure that at this scale all the
structure of the kernel at higher momenta is irrelevant to $\Delta E$.

\subsection{Fixing of the parameters after ghost subtraction}

As noted in section \ref{II}, the equation of state at zero
temperature for nuclear matter depends only on the dimensionless
quantities $C^2_s$ and $C_v^2$, that now become
\begin{equation}
C_s^2 = \wh{g}_s^2 \, \frac{M^2}{\wh{m}_s^2}\,,\qquad
C_v^2 = \wh{g}_v^2 \, \frac{M^2}{\wh{m}_v^2}\,.
\label{CsCv}
\end{equation}
Fixing the saturation density and binding energy to their observed
values yields, of course, the same numerical values for $C_s^2$ and
$C_v^2$ as in the original theory. After this is done, all static
properties of nuclear matter are determined and thus they are
insensitive to the ghost subtraction. Therefore, at leading order in
the $1/N$ expansion, to see any effect one should study either
dynamical nuclear matter properties as done in ref.~\cite{Ta91} or
finite nuclei as we do here.

It is remarkable that if all the parameters of the model were to be
fixed exclusively by a set of nuclear structure properties, the ghost
subtracted and the original theories would be indistinguishable
regarding any other static nuclear prediction, because bosonic and
fermionic equations of motion have the same form in both
theories. They would differ however far from the zero four momentum
region where the truncation of the ghost kernels $\Delta K(p^2)$ at
order $p^2$ is no longer justified. In practice, the predictions will
change after ghost removal because the $\omega$-meson mass is quite large
and is one of the observables to be used in the fixing of the
parameters.

To fix the parameters of the theory we choose the same observables as 
in section \ref{II}. Let us consider first the vector meson
parameters $\wh{m}_v$ and $\wh{g}_v$. We proceed as follows:

1. We choose a trial value for $g_v$ (the zero-momentum coupling
constant of the original theory). This value and the known physical
values of the $\omega$-meson and nucleon masses, $m_\omega$ and $M$
respectively, determines $m_v$ (the zero-momentum mass of the original
theory), namely
\begin{equation}
m_v^2 = m_\omega^2 + \frac{\gamma \, g_v^2}{8 \, \pi^2}\, M^2\,
\left\{ \frac{4}{3} + \frac{5}{9}\, \frac{m_\omega^2}{M^2}
- \frac{2}{3} \left(2 + \frac{m_\omega^2}{M^2}\right)
\sqrt{\frac{4 \, M^2}{m_\omega^2} - 1 } \,
\arcsin\left(\frac{m_\omega}{2 \, M}\right)\right\} \,.
\end{equation}
(This, as well as the formulas given below, can be deduced from those
in the appendix.)

2. $g_v$ and $m_v$ provide the values of the tachyonic
parameters $R_G^v$ and $M_G^v$. They are given by
\begin{eqnarray}
M_G^v &=& \frac{2M}{\sqrt{\kappa_v^2-1}} \nonumber \\
\frac{1}{R_G^v} &=& -1 + \frac{\gamma \, g_v^2}{24\,\pi^2}
\left\{ \left(\frac{\kappa_v^3}{4} + \frac{3}{4\,\kappa_v}\right)
 \log \frac{\kappa_v + 1}{\kappa_v - 1} - \frac{\kappa_v^2}{2} -
 \frac{1}{6}\right\}\,,
\label{RGv}
\end{eqnarray}
where the quantity $\kappa_v$ is the real solution of the following
equation (there is an imaginary solution which corresponds to the
$\omega$-meson pole)
\begin{equation}
1 + \frac{m_v^2}{4 \, M^2} \, (\kappa_v^2-1)
+\frac{\gamma \, g_v^2}{24\,\pi^2}
\left\{ \left(\frac{\kappa_v^3}{2} - \frac{3\, \kappa_v}{2}\right)
 \log \frac{\kappa_v + 1}{\kappa_v - 1} - \kappa_v^2 +
\frac{8}{3}\right\} =0 \,.
\label{kappav}
\end{equation}

3. Known $g_v$, $m_v$, $M_G^v$ and $R_G^v$, the values of $\wh{m}_v$ and
$\wh{g}_v$ are obtained from eqs.~(\ref{mg}). They are then inserted
in eqs.~(\ref{CsCv}) to yield $C^2_v$. If necessary, the initial trial
value of $g_v$ should be readjusted so that the value of $C^2_v$ so
obtained coincides with that determined by the saturation properties
of nuclear matter.

The procedure to fix the parameters $m_s$ and $g_s$ is similar but
slightly simpler since the physical mass of the scalar meson
$m_\sigma$ is not used in the fit. Some trial values for $m_s$ and
$g_s$ are proposed. This allows to compute $M_G^s$ and $R_G^s$ by
means of the formulas
\begin{eqnarray}
M_G^s &=& \frac{2M}{\sqrt{\kappa_s^2-1}} \nonumber \\
\frac{1}{R_G^s} &=& -1 - 
\frac{\gamma \, g_s^2}{16\,\pi^2}
\left\{ \left(\frac{\kappa_s^3}{2} \, - \frac{3\,\kappa_s}{2}\right) 
\log \frac{\kappa_s + 1}{\kappa_s - 1}
- \kappa_s^2 + \frac{8}{3}\right\}\,,
\label{RGs}
\end{eqnarray}
where $\kappa_s$ is the real solution of 
\begin{equation}
1 + \frac{m_s^2}{4 \, M^2} \, (\kappa_s^2-1)
- \frac{\gamma \, g_s^2}{16\,\pi^2}
\left\{ \kappa_s^3 \, \log \frac{\kappa_s + 1}{\kappa_s - 1}
- 2 \, \kappa_s^2 - \frac{2}{3}\right\} = 0\,.
\label{kappas}
\end{equation}
One can then compute $\wh{m}_s$ and $\wh{g}_s$ and thus $C_s^2$ and
the mean quadratic charge radius of $^{40}$Ca. The initial values of
$m_s$ and $g_s$ should be adjusted to reproduce these two quantities.
We will refer to the set of masses and coupling constants so obtained
as the {\em no-ghost scheme} parameters.

\section{Numerical results and discussion}
\label{IV}

As explained in section~\ref{II}, the parameters of the theory are
fitted to five observables. For the latter we take the following
numerical values: $M=939$~MeV, $m_\omega=783$~MeV, $B/A=15.75$~MeV,
$k_F=1.3$~fm$^{-1}$ and $3.82$~fm for the mean quadratic charge radius
of $^{40}$Ca.

If the Dirac sea is not included at all, the numerical values that we
find for the nuclear matter combinations $C_s^2$ and $C_v^2$ are
\begin{equation}
C_s^2 = 357.7\,, \qquad C_v^2 = 274.1 
\end{equation}
The corresponding Lagrangian parameters are shown in
table~\ref{t-par-1}. There we also show $m_\sigma$ and $m_\omega$ that
correspond to the position of the poles in the propagators after
including the one-loop meson self energy. They are an output of the
calculation and are given for illustration purposes.

When the Dirac sea is included, nuclear matter properties fix the
following values
\begin{equation}
C_s^2 = 227.8\,, \qquad C_v^2 = 147.5
\end{equation}
Note that in nuclear matter only the zeroth order $E_0^{\rm sea}$ is
needed in the gradient expansion of the sea energy, since the meson
fields are constant. The (zero momentum renormalized) Lagrangian meson
masses $m_{s,v}$ and coupling constants $g_{s,v}$ are shown in
table~\ref{t-par-1} in various schemes, namely, $\omega$-shell,
no-ghost and naive schemes, previously defined. The scalar meson
parameters differ if the Dirac sea energy is included at zeroth order
or at all orders (in practice zeroth plus second order) in the
gradient expansion. For the sake of completeness, both possibilities
are shown in the table. The numbers in brackets in the no-ghost scheme
are the zero-momentum parameters of the effective action,
$\wh{m}_{s,v}$ and $\wh{g}_{s,v}$ (in the other schemes they coincide
with the Lagrangian parameters). Again $m_\sigma$ and $m_\omega$ refer
to the scalar and vector propagator-pole masses after including the
one fermion loop self energy for each set of Lagrangian parameters.
Table~\ref{t-par-2} shows the ghost masses and residues corresponding
to the zero-momentum renormalized propagators. The no-ghost scheme
parameters have been used.

The binding energies per nucleon (without center of mass corrections)
and mean quadratic charge radii (without convolution with the nucleon
form factor) of several closed-shell nuclei are shown in
tables~\ref{t-dat-1} and \ref{t-dat-2} for the $\omega$-shell and for
the naive and no-ghost schemes (these two schemes give the same
numbers), as well as for the case of not including the Dirac sea.
The experimental data are taken from refs.~\cite{Ja74,Va72,Wa88a}.

From table~\ref{t-par-1} it follows that the zero-momentum vector meson
mass $m_v$ in the $\omega$-shell scheme is considerably larger than the
physical mass. This is somewhat unexpected. Let us recall that the
naive treatment, which neglects the meson self energy, is the most
used in practice. It has been known for a long time~\cite{Pe86,FH88}
that the $\omega$-shell scheme is, as a matter of principle, the
correct procedure but on the basis of rough estimates it was assumed
that neglecting the meson self energy would be a good approximation
for the meson mass. We find here that this is not so.

Regarding the consequences of removing the ghost, we find in
table~\ref{t-par-1} that the effective parameters $\wh{m}_{s,v}$ and
$\wh{g}_{s,v}$ in the no-ghost scheme are similar, within a few per
thousand, to those of the naive scheme. This similarity reflects in
turn on the predicted nuclear properties: the results shown in
tables~\ref{t-dat-1} and \ref{t-dat-2} for the no-ghost scheme
coincide, within the indicated precision, with those of the naive
scheme (not shown in the table). It is amazing that the outcoming
parameters from such a sophisticated fitting procedure, namely the
no-ghost scheme, resemble so much the parameters corresponding to the
naive treatment. We believe this result to be rather remarkable for it
justifies a posteriori the nowadays traditional calculations made with
the naive scheme.

\begin{figure}
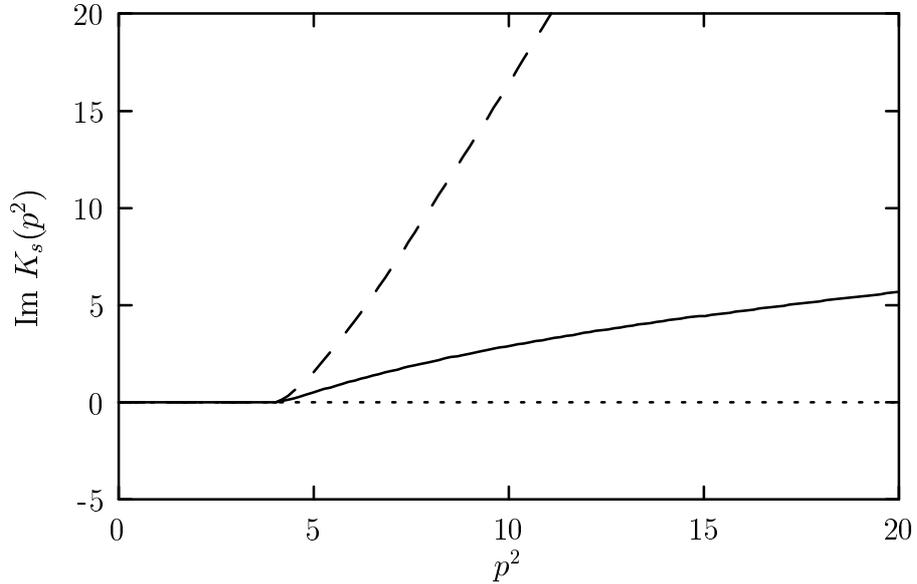

\caption{
Imaginary part of the inverse scalar meson propagator
$K_s(p^2)$. Units and meaning of the lines as in fig.~1.
}
\label{f-imaginary}
\end{figure}

The above observation is equivalent to the fact that the zero-momentum
masses, $\wh{m}_{s,v}$, and the propagator-pole masses
$m_{\sigma,\omega}$ are very similar in the no-ghost scheme. This
implies that the effect of removing the ghosts cancels to a large
extent with that introduced by the meson self energies. Note that
separately the two effects are not small; as was noted above $m_v$ is
much larger than $m_\omega$ in the $\omega$-shell scheme. To interpret
this result, it will be convenient to recall the structure of the
meson propagators.  In the leading $1/N$ approximation, there are
three kinds of states that can be created on the vacuum by the meson
fields. Correspondingly, the spectral density functions $\rho(q^2)$
have support in three clearly separated regions, namely, at the ghost
mass squared (in the Euclidean region), at the physical meson mass
squared, and above the $N\overline{N}$ pair production threshold
$(2M)^2$ (in the time-like region). The full meson propagator is
obtained by convolution of the spectral density function with the
massless propagator $(q^2+i\epsilon)^{-1}$ as follows from the
K\"alle\'en-Lehmann representation, eq.~(\ref{KL}). The large
cancelation found after removing the ghosts leads to the conclusion
that, in the zero-momentum region, most of the correction induced by
the fermion loop on the meson propagators, and thereby on the
quadratic kernels $K(p^2)$, is spurious since it is due to unphysical
ghost states rather than to virtual $N\overline{N}$ pairs. This can
also be seen from figs.~\ref{f-real} and \ref{f-imaginary}. There, we
represent the real and imaginary parts of $K_s(p^2)$ respectively, in
three cases, namely, before ghost elimination, after ghost elimination
and the free inverse propagator. In all three cases the slope of the
real part at zero momentum is equal to one and the no-ghost (sea 2nd)
set of parameters from table~\ref{t-par-1} has been used. We note the
strong resemblance of the free propagator and the ghost-free
propagator below threshold. A similar result is obtained for the
vector meson.

One may wonder how these conclusions reflect on the sea energy. Given
that we have found that most of the fermion loop is spurious in the
meson self energy it seems necessary to revise the sea energy as well
since it has the same origin. Technically, no such problem appears in
our treatment. Indeed the ghost is found in the fermion loop attached
to two meson external legs, i.e., terms quadratic in the fields.
However, the sea energy used, namely, $E_0^{\rm sea}+E_2^{\rm sea}$,
does not contain such terms. Quadratic terms would correspond to a
mass term in $E_0^{\rm sea}$ and a kinetic energy term in $E_2^{\rm
sea}$, but they are absent from the sea energy due to the
zero-momentum renormalization prescription used. On the other hand,
terms with more than two gradients were found to be
negligible~\cite{Ca96a}. Nevertheless, there still exists the
possibility of ghost-like contributions in vertex functions
corresponding to three or more mesons, similar to the spurious
contributions existing in the two-point function. In this case the
total sea energy would have to be reconsidered. The physically
acceptable dispersion relations for three or more fields have been
much less studied in the literature hence no answer can be given to
this possibility at present.

\section{Summary and conclusions}
\label{V}

We summarize our points. In the present paper, we have studied the
consequences of eliminating the vacuum instabilities which take place
in the $\sigma$-$\omega$ model. This has been done using Redmond's
prescription which imposes the validity of the K\"all\'en-Lehmann
representation for the two-point Green's functions. We have discussed
possible interpretations to such method and have given plausibility
arguments to regard Redmond's method as a non perturbative and non
local modification of the starting Lagrangian. 

Numerically we have found that, contrary to the naive expectation, the
effect of including fermionic loop corrections to the mesonic
propagators ($\omega$-shell scheme) is not small. However, it largely
cancels with that of removing the unphysical Landau poles. A priori,
this is a rather unexpected result which in fact seems to justify
previous calculations carried out in the literature using a naive
scheme. Actually, as compared to that scheme and after proper
readjustment of the parameters to selected nuclear matter and finite
nuclei properties, the numerical effect becomes rather moderate on
nuclear observables. The two schemes, naive and no-ghost, are
completely different beyond the zero four momentum region, however,
and for instance predict different values for the vector meson mass.

Therefore it seems that in this model most of the fermionic loop
contribution to the meson self energy is spurious. The inclusion of
the fermionic loop in the meson propagator can only be regarded as an
improvement if the Landau ghost problem is dealt with simultaneously.
We have seen that the presence of Landau ghosts does not reflect on
the sea energy but it is not known whether there are other
spurious ghost-like contributions coming from three or higher point
vertex functions induced by the fermionic loop.

\section*{Acknowledgments}
This work is supported in part by funds provided by the U.S.
Department of Energy (D.O.E.) under cooperative research agreement
\#DF-FC02-94ER40818, Spanish DGICYT grant no. PB92-0927 and Junta de
Andaluc\'{i}a grant no. FQM0225. One of us (J.C.)  acknowledges the
Spanish Ministerio de Educaci\'on y Ciencia for a research grant.

\appendix{}\section{}

\subsection{Meson self energies in the leading $1/N$ expansion}
As stated in the main text, the leading order in the  $1/N$ expansion
($N$ being the number of nucleon species) is achieved by considering
1-fermion loop  and zero-boson loop Feynman graphs in the effective action.
This corresponds to compute the meson self energies at 1-loop
approximation.

For the $\sigma$-meson, the bare self energy in terms of the Lagrangian
coupling constant is obtained as
\begin{eqnarray}
\Pi_{B,s}(p^2; M, \xi, \varepsilon) & = &
- i\, \xi^{2 \varepsilon} \, \int \frac{d^{4 - 2 \varepsilon} k}
{(2 \pi)^{4 - 2 \varepsilon}} \, \mbox{Tr }\Biggl\{ \Biggr.
\frac{i}{ \dslash{p} + \dslash{k} - M + i \epsilon}
\, i g_s \, \times \nonumber\\
& & \quad \times
\frac{i}{ \dslash{k} - M + i \epsilon}
\, i g_s \Biggl. \Biggr\}\,.
\end{eqnarray}
Imposing zero-momentum renormalization we get
\begin{eqnarray}
\Pi_s(p^2) & = & \left[ \Pi_{B,s}(p^2) - \Pi_{B,s}(0) - \Pi_{B,s}'(0)
\, p^2 \right] \nonumber \\
           & = & - \frac{g_s^2 N}{4 \pi^2} \Biggl\{ \Biggr.
\left(2 M^2 - \frac{1}{2} p^2\right) \, I_a\left(\frac{p^2}{M^2}\right) +
 \frac{p^2}{3}
\Biggl. \Biggr\}\,,
\end{eqnarray}
where the function $I_a(y)$ is defined as
\begin{eqnarray}
I_a(y) & = &
\int_0^1 dx \, \log \left[1 - y x (1-x) - i \epsilon\right]
\nonumber \\
       & = &\left\{
\begin{array}{ll}
\sqrt{1 - \frac{4}{y}} \, \log \frac{\sqrt{1 - \frac{4}{y}}\, +
1}{\sqrt{1 -\frac{4}{y}}\, - 1} 
- 2 &
\phantom{0<}  y < 0 \\
2 \sqrt{\frac{4}{y} - 1}\, \arcsin\left(\frac{\sqrt{y}}{2}\right) - 2
&
0<  y < 4 \\
\sqrt{1 - \frac{4}{y}} \, \log
\frac{1 + \sqrt{1 - \frac{4}{y}}}{1 - \sqrt{1 -\frac{4}{y}}} 
- 2 - i \sqrt{1 - \frac{4}{y}} \, \pi \quad
&
4 < y \,.
\nonumber \\
\end{array}
\right.
\end{eqnarray}

The $\omega$-meson self energy is obtained in a similar way
but taking care of its Lorentz structure, 
\begin{eqnarray}
\Pi_{B,v}^{ \mu \nu}(p^2; M, \varepsilon, \xi) & = &
- i\, \xi^{2 \varepsilon} \, \int \frac{d^{4 - 2 \varepsilon} k}
{(2 \pi)^{4 - 2 \varepsilon}} \, \mbox{Tr } \Biggl\{ \Biggr.
\frac{i}{ \dslash{p} + \dslash{k} - M + i \epsilon}
\, (-i g_v) \gamma^\mu \, \times \nonumber\\
& & \quad \times
\frac{i}{ \dslash{k} - M + i \epsilon}
\, (-i g_v) \gamma^\nu \Biggl. \Biggr\}\,,
\end{eqnarray}
which is highly simplified by baryonic current conservation,
\begin{equation}
\Pi_{B,v}^{\mu \nu}(p^2)= \left(-g^{\mu \nu} + 
\frac{p^\mu p^\nu}{p^2} \right)\Pi_{B,v}(p^2)\,.
\end{equation}
The explicit expression of the $\omega$-meson self energy renormalized at
zero-momentum is
\begin{eqnarray}
\Pi_v(p^2) & = & \Pi_{B,v}(p^2) - \Pi_{B,v}(0)
-  \Pi_{B,v}'(0) \, p^2
\nonumber \\
& = &
- \frac{N\, g_v^2}{12 \pi^2}
\Biggl\{\Biggr.
(2 M^2 + p^2) I_a\left(\frac{p^2}{M^2}\right)
+ \frac{p^2}{3}
\Biggl.\Biggr\}\,.
\end{eqnarray}

\subsection{Poles and residues}

The relation between the Lagrangian mass of a boson, $m$, and its
physical mass, $m_{\rm sh}$, is given by
\begin{equation}
 m_{\rm sh}^2 - m^2 - \Pi(m_{\rm sh}^2) = 0\,,
\end{equation}
where the self energy, $\Pi(p^2)$, is assumed to be renormalized at zero
momentum. 
From the expression of its self energy given above we find that
the $\sigma$-meson physical pole, $m_\sigma$, can be obtained in terms of the
Lagrangian parameters $g_s$ and $m_s$ by solving the transcendental
equation 
\begin{equation}
m_s^2 = m_\sigma^2 -  \frac{N\, g_s^2}{4 \pi^2}\, M^2 \,
\left[
4 - \frac{4}{3} \, \frac{m_\sigma^2}{M^2} +
\left(- 4 + \frac{m_\sigma^2}{M^2}\right)\,
\sqrt{\frac{4 M^2}{m_\sigma^2} - 1} \, \arcsin \frac{m_\sigma}{2 M}
\right] \,.
\label{E-OS-ms}
\end{equation}
Similarly, the equation to solve for the $\omega$ particle is
\begin{equation}
m_v^2 = m_\omega^2  +  \frac{N\, g_v^2}{4 \pi^2}\,M^2 \,
\left[
\frac{4}{3} + \frac{5}{9} \, \frac{m_\omega^2}{M^2}
- \frac{2}{3} \left(2 + \frac{m_\omega^2}{M^2}\right)\,
\sqrt{\frac{4 M^2}{m_\omega^2} - 1} \, \arcsin \frac{m_\omega}{2 M}
\right]\,.
\label{E-OS-mv}
\end{equation}
It is interesting to note that sometimes the combination $C^2 = M^2
g^2/m^2$ is taken to be fixed by nuclear matter properties. This
allows one to write the Lagrangian coupling constant, $g$, as a
function of $C$ and the Lagrangian mass, $m$. Inserting the omega
version of this expression into the previous equation permits to solve
the Lagrangian mass in terms of $C_v$ and the physical $\omega$
mass, $m_\omega$.  If the eqs.~(\ref{E-OS-ms}-\ref{E-OS-mv}) are
conveniently extended to the $m_{\rm sh}$-complex plane they can be
used to obtain the Landau ghost masses as well (better expressions for
numerical calculation are found in the main text in eqs.~(\ref{kappas}) and
(\ref{kappav})).

Once a Landau pole has been computed, the value of 
its zero-momentum residue, $-R_G$, is easily obtained as
\begin{equation}
- R_G = 1 - \Pi'(-M_G^2)\,.
\end{equation}
The particular expressions of this equation for the $\sigma$ and $\omega$
meson are given in eqs.~(\ref{RGs}) and (\ref{RGv}) respectively.

\begin{table}
$$
\begin{array}{*{8}{|c}|}
\cline{3-8}
\multicolumn{2}{c|}{ }& g_s & m_s & m_\sigma &
	g_v & m_v & m_\omega \\
\hline
\multicolumn{2}{|c|}{\mbox{no sea}}
& 9.062 & 449.9 & 439.8 & 13.81 & 783\phantom{.1} & 673.6 \\
\hline
& \omega\mbox{-shell} &
6.153 & 382.8 & 379.8 & 11.78 & 910.8 & 783\phantom{.1} \\
\mbox{sea 0th} & \mbox{no-ghost} & 5.996 & 370.9 & 368.3 & 
14.86 & 978.6 & 783\phantom{.1} \\
&  & (5.928) & (368.8) &  & (10.17) & (786.1) &  \\
& \mbox{naive} &
5.922 & 368.4 & 365.9 & 10.13 & 783\phantom{.1} & 711.2\\ 
\hline
& \omega\mbox{-shell} &
6.846 & 425.9 & 420.3 & 11.78 & 910.8 & 783\phantom{.1} \\
\mbox{sea 2nd} & \mbox{no-ghost} &
6.664 & 410.8 & 406.5 & 14.86 & 978.6 & 783\phantom{.1} \\
& & (6.544) & (407.1) &  & (10.17) & (786.1) &  \\
& \mbox{naive} &
6.536 & 406.6 & 402.6 & 10.13 & 783\phantom{.1} & 711.2 \\ 
\hline
\end{array}
$$
\caption{Zero momentum renormalized Lagrangian parameters in several
schemes. Masses are in MeV. The meaning of the labels no sea, sea 0th,
sea 2nd, $\omega$-shell and no-ghost are as in table III. The naive
scheme corresponds to not including the meson self energy. The numbers
in brackets are the zero-momentum parameters of the effective action
for the no-ghost scheme, $\wh{m}_{s,v}$ and $\wh{g}_{s,v}$. In all
cases, $m_\sigma$ and $m_\omega$ stand for the poles in the meson
propagators after including the one fermion loop self energy and using
the corresponding Lagrangian parameters. Note that by construction the
vector meson parameters coincide in the sea 0th and sea 2nd cases.}
\label{t-par-1}
\end{table}

\begin{table}
$$
\begin{array}{*{5}{|c}|}
\cline{2-5}
\multicolumn{1}{c|}{ }& R_G^s & M_G^s & R_G^v & M_G^v \\
\hline
\mbox{sea 0th} & 1.748 & 4605 & 0.6090 & 1457 \\
\mbox{sea 2nd} & 1.584 & 3863 & 0.6090 & 1457 \\
\hline
\end{array}
$$
\caption{Residue (up to a sign) and mass (in MeV) of the ghosts in the
zero-momentum renormalized meson propagators using the no-ghost sets
of Lagrangian parameters in table I.}
\label{t-par-2}
\end{table}

\begin{table}[hbt]
\begin{center}
$$
\begin{array}{|c|r@{.}l|*{2}{r@{.}l|}*{2}{r@{.}l|}r@{.}l|}
\cline{2-13}
\multicolumn{1}{c|}{} &
\multicolumn{12}{|c|}{\mbox{B/A (MeV)}}\\
\cline{2-13}
\multicolumn{1}{c|}{ }
        & \multicolumn{2}{|c|}{\mbox{no sea}}
        & \multicolumn{4}{|c|}{\mbox{sea 0th}}
        & \multicolumn{4}{|c|}{\mbox{sea 2nd}}
        & \multicolumn{2}{|c|}{\mbox{exp.}}
\\ \cline{1-1} \cline{4-11}
^A_Z\mbox{X} & \multicolumn{2}{|c|}{ }
        & \multicolumn{2}{|c|}{\mbox{no-ghost}}
        & \multicolumn{2}{|c|}{\omega\mbox{-shell}}
        & \multicolumn{2}{|c|}{\mbox{no-ghost}}
        & \multicolumn{2}{|c|}{\omega\mbox{-shell}}
        & \multicolumn{2}{|c|}{}
\\ \hline
^{40}_{20}\mbox{Ca}  & 6&28   &
		  6&00   & 6&10   &
		  6&33   & 6&43   & 
                  8&55                   \\
^{56}_{28}\mbox{Ni}  & 7&24   & 
                  6&51   & 6&60 &
                  6&80   & 6&90 &
                  8&64\\
^{90}_{40}\mbox{Zr}  & 8&36   &
		  7&99   & 8&07 &
		  8&22   & 8&30 &
                  8&71\\
^{132}_{50}\mbox{Sn} & 8&81   &
		  8&43   & 8&50   &
		  8&62   & 8&69   &
                  8&36   \\
^{208}_{\phantom{2}82}\mbox{Pb} & 9&84 &
			     9&55 & 9&61 &
	       		     9&70 & 9&76 &
                             7&87\\
\hline
\end{array}
$$
\caption{Binding energy per nucleon of some closed-shell nuclei
computed in several ways: not including the Dirac sea in the parameter
fixing (no-sea), including the Dirac sea at lowest order in a gradient
expansion (sea 0th), including the Dirac sea at all orders (sea 2nd)
and the experimental values (exp.). The entry $\omega$-shell
corresponds to use the set of parameters that reproduce the
$\omega$-meson mass after including the meson self energy. The entry
no-ghost corresponds to the parameters obtained by applying Redmond's
prescription.}
\label{t-dat-1}
\end{center}
\end{table}

\begin{table}[hbt]
\begin{center}
$$
\begin{array}{|c|r@{.}l|*{2}{r@{.}l|}*{2}{r@{.}l|}r@{.}l|}
\cline{2-13}
\multicolumn{1}{c|}{} &
\multicolumn{12}{|c|}{\mbox{m.q.c.r. (fm)}}\\
\cline{2-13}
\multicolumn{1}{c|}{ }
        & \multicolumn{2}{|c|}{\mbox{no sea}}
        & \multicolumn{4}{|c|}{\mbox{sea 0th}}
        & \multicolumn{4}{|c|}{\mbox{sea 2nd}}
        & \multicolumn{2}{|c|}{\mbox{exp.}}
\\ \cline{1-1} \cline{4-11}
^A_Z\mbox{X} & \multicolumn{2}{|c|}{ }
        & \multicolumn{2}{|c|}{\mbox{no-ghost}}
        & \multicolumn{2}{|c|}{\omega\mbox{-shell}}
        & \multicolumn{2}{|c|}{\mbox{no-ghost}}
        & \multicolumn{2}{|c|}{\omega\mbox{-shell}}
        & \multicolumn{2}{|c|}{}
\\ \hline
^{40}_{20}\mbox{Ca}^*   & 3&48   &
		    3&48     & 3&48   &
		    3&48     & 3&48   & 
                    3&48                   \\
^{56}_{28}\mbox{Ni}  & 3&72   & 
                  3&79   & 3&79 &
                  3&79   & 6&80 &
                  \multicolumn{2}{|c|}{}\\
^{90}_{40}\mbox{Zr}  & 4&22   &
		  4&23   & 4&24 &
		  4&25   & 4&25 &
                  4&27\\
^{132}_{50}\mbox{Sn} & 4&60   &
		  4&66   & 4&66   &
		  4&68   & 4&68   &
                  \multicolumn{2}{|c|}{}\\
^{208}_{\phantom{2}82}\mbox{Pb} &5&35   &
              		    5&39   & 5&39   &
		            5&41   & 5&41   &
                            5&50   \\
\hline
\end{array}
$$
\caption{Mean quadratic charge radii of several closed-shell
nuclei. Meaning of the labels and experimental values as in table
III.}
\label{t-dat-2}
\end{center}
\end{table}

\end{document}